\documentclass[aip,jcp,preprint,amssymb,superscriptaddress]{revtex4-1}
\usepackage{amsmath}
\usepackage{latexsym}
\usepackage{mathrsfs}
\usepackage{graphicx}
\usepackage{dcolumn}
\usepackage{epsfig}
\usepackage{graphics}
\usepackage{bm}
\usepackage{units,nicefrac}

\begin{document}

\title{
A Method for Analyzing
the Non-Stationary Nucleation\\ and Overall Transition Kinetics. \\
A Case of Water }

\author{
Anatolii V. Mokshin }\email{anatolii.mokshin@mail.ru}
\affiliation{Kazan Federal University, Kremlevskaya Street 18,
420000 Kazan,  Russia }
\author{Bulat N. Galimzyanov} \affiliation{Kazan Federal University,
Kremlevskaya Street 18, 420000 Kazan,  Russia}

\date{\today}

\begin{abstract}
We present the statistical method as a direct extension of the
mean first-passage time concept to the analysis of molecular
dynamics simulation data of a phase transformation. According to
the method, the mean first-passage time trajectories for the first
$(i=1)$ as well as for the subsequent ($i=2$, $3$, $4$, $\ldots$)
nucleation events should be extracted, that allows one to
calculate the time-dependent nucleation rate, the critical value
of the order parameter (the critical size), the waiting times for
the nucleation events and the growth law of the nuclei -- i.e. all
the terms, which are usually necessary to characterize the overall
transition kinetics. There are no restrictions in the application
of the method by the specific thermodynamic regions; and the
nucleation rate parameters are extracted according to their basic
definitions. The method differs from the Wedekind-Bartell scheme
and its modification [A.V.~Mokshin, B.N. Galimzyanov, J. Phys.
Chem. B {\bf 116}, 11959 (2012)], where the passage-times for the
first (largest) nucleus are evaluated only and where the average
waiting time for the first nucleation event is accessible instead
of the true steady-state nucleation time scale. We demonstrate an
efficiency of the method by its application to the analysis of the
vapor-to-liquid transition kinetics in water at the different
temperatures. The nucleation rate/time characteristics and the
droplet growth parameters are computed on the basis of the
coarse-grained molecular dynamics simulation data.

\end{abstract}


\maketitle

\section{Introduction}

The numerical methods of the molecular (particle) dynamics
simulations are appearing to be useful tool at the study of the
phase transitions~\cite{Frenkel_book_1996}. The detailed
information about the instantaneous molecular configurations as
well as the trajectories of `molecules', that is accessible due to
these simulations, prompts to revise the results of the
traditional experiments, and to develop the corresponding
theoretical descriptions. Hence, there is a necessity for such the
methods, which allows one to compute thermodynamic and kinetic
characteristics of the phase transitions within the molecular
dynamics (MD) simulation data. Importantly, although the different
phase transformations (crystallization, evaporation, condensation
etc.) are characterized by quite different physical processes,
there are similar regimes in their passing with
time~\cite{Frenkel_book_1946}. Therefore, the methods developed
originally for the computation of the rate characteristics for the
some concrete transition (say, crystallization) can be extended
with equal success to any first-order phase transition.

One of the most important properties of the phase transition
kinetics is the nucleation rate $J(t)$. This quantity defines the
number of the \textit{overcritical} nuclei formed in the
mother-phase per unit time and unite volume, which are able to
grow to the new bulk
phase~\cite{Kelton_review_1991,Deben_book_1996,Skripov_Metastable_liquids_1974}.
Although the nucleation rate is the time-dependent, the some
methods adapted specially for its computation within the MD
simulations are restricted by the consideration of the
steady-state nucleation regime, where $J(t) = J_s$. One can
mention thereupon the Yasuoka-Matsumoto
method~\cite{Yasuoka/Matsumoto_JCP_1998}, the Volkov
method~\cite{Volkov_PRE_2002} and the method of survival
probability~\cite{Saika-Voivod_JCP_2009,Chushak/Bartell}, which
are aimed for the computation of $J_s$. Another scheme used
frequently for the analysis of the MD simulation results was
suggested in Refs.~\cite{Wedekind_JCP_2006,Bartell_JCP_2006} and
is known as the Wedekind-Bartell (WB) scheme. It is based on the
computation of the average time necessary for the earliest nucleus
to reach the different sizes $n$ for the first time, that is, in
fact, the mean first-passage time for the parameter $n$ (see
Refs.~\cite{Hanggi_RMP_1990,Redner_book_2001}). According to
Refs.~\cite{Wedekind_JCP_2006,Bartell_JCP_2006}, the critical size
of the nucleus, $n_c$, and the steady-state nucleation rate $J_s$
can be defined via the resulting mean first-passage time (MFPT)
curve generated from the multiple MD simulations.

In this work, we show that the WB-scheme as well as its extended
version (see Ref.~\cite{Mokshin/Galimzyanov_JPCB_2012}) do not
provide the proper steady-state nucleation rate. Therewith, we
present a novel method of statistical analysis, which is based on
consideration of the MFPT trajectories defined for the whole
succession of the nucleation events. In particular, the method can
be applied to extract the time-dependent nucleation rate, the
average waiting times of the nucleation events, the induction
time, the growth laws and, thereby, to quantify the overall
transition kinetics. We demonstrate the application of the method
to evaluation of the non-stationary droplet nucleation and of the
overall vapor-to-liquid transition in water.

\section{Description of the Method \label{Sec: MFPT}}

Let us assume that the order parameter $n$ characterizes
transition of the system from metastable state to stable one,
where both states are separated in a free energy landscape $\Delta
G(n)$ by a barrier $\Delta G_{n_c}$. Although the stochastic
features are pronounced in the behavior of the order parameter $n$
during the transition~\cite{Hanggi_RMP_1990}, the time scale
$\tau_s$ (or by the rate $1/\tau_s$) of the barrier crossing could
be defined in \textit{statistical
sense}~\cite{Skripov_Metastable_liquids_1974}. Practically, the
estimation of the term $\tau_s$ can be carried out by means of the
method of inverted averaging, which is a basis of the MFPT
concept~\cite{Redner_book_2001}. Let us suppose that we are able
to observe the temporal evolution of the order parameter
$n_{\alpha}(t)$ in an $\alpha$th experimental run within the
ensemble of $M$ experiments, $\alpha \in [1,\;M]$. From every
known trajectory $n_{\alpha}(t)$, one can define the times of the
first appearance of the order parameter with the concrete value,
i.e. $\tau_{\alpha}(n)$. Then, the mean first-passage times
$\bar{\tau}(n)$ comes directly as an average of $\tau_{\alpha}(n)$
over all the experiments,
\begin{equation} \label{eq: MFPT_single}
\bar{\tau}(n) = \frac{1}{M} \sum_{\alpha=1}^{M}\tau_{\alpha}(n).
\end{equation}
The position of the first inflection point in $\bar{\tau}(n)$
corresponds to the critical value $n_c$ -- that is a location of
the free energy barrier $\Delta G_{n_c}$. Moreover, the quantity
$\bar{\tau}(n_c)$ defines the most probable time scale of the
appearance of the critical value $n_c$ within the statistics of
the $M$ experimental pathways for $n_{\alpha}(t)$. This is a
general inference of the MFPT
concept~\cite{Redner_book_2001,Pontryagin_JETP_1933,Hanggi_RMP_1990}.

This concept can be \emph{directly} extended to analyzing the
nucleation-growth processes. Nucleation is a typical activated
process, which represents an initial stage of the phase transition
in many-particle system. Moreover, the transition proceeds through
the consecutive nucleation events. The following temporal
characteristics appears naturally: the waiting time of the first
nucleation event $\tau_1$ and the nucleation rate $J(t)$. The
quantity $\tau_1$ defines the time scale necessary to form the
first (earliest) nucleus of the critical
size~\cite{Koishi_JCP_2003} and, thereby, represents a lifetime of
the metastable state. Therefore, the quantity $\tau_1$ corresponds
to a time scale, for which the nucleation rate takes the first
\textit{detectable} nonzero value, i.e. $J(\tau_1) =
1/[V(t)\tau_1]$, where  $V$ is the volume of a system. Note that
the regime of the steady-state nucleation with the rate $J(t)=J_s$
can be established in the system after a lapse of the time
associated with the so-called induction time
$\tau_{ind}$~\cite{Fokin_Nucl2005}.

Let us associate the order parameter with the nucleus size
$n_{\alpha i}(t)$, which is defined as the number of molecules in
the nucleus at the time $t$. The term $\alpha$ denotes the index
of simulation run, whereas the order number of the nucleation
event $i$ indicates that the $i$th nucleus appears during the
$\alpha$th simulation run. For each $\alpha$th simulation, we
extract the pathway with the $i$th nucleation event --  $n_{\alpha
i}(t)$, $i=1,2,3,...$  -- at the condition that the configurations
for the previous pathways [$n_{\alpha j}(t)$, $j=1,2,...,(i-1)$]
are ignored. According to Eq.~(\ref{eq: MFPT_single}), we generate
within the extracted data for $\tau_{\alpha i}(n)$ the whole set
of
\begin{equation} \label{eq: MFPT_multiple}
\bar{\tau}_i(n) = \frac{1}{M} \sum_{\alpha=1}^{M}\tau_{\alpha
i}(n), \ \ \ i=1,\;2,\;3,\ldots,
\end{equation}
where the quantity $\bar{\tau}_i(n)$ at the fixed $i$ represents
the mean first-passage times for the size $n$, that is attained by
the $i$th nucleus. Hence, the quantity $\bar{\tau}_1(n)$
corresponds to MFPT's for the earliest nucleus, then
$\bar{\tau}_2(n)$ is the MFPT's for the second nucleus, etc.

What information can be recovered from $\bar{\tau}_i(n)$?

(i) According to the MFPT concept, the first inflection point on
each the curve $\bar{\tau}_i(n)$, where the term $\partial
\bar{\tau}_i(n)/\partial n$ has a maximum, will determine the
critical size $n_c^{(i)}$ of the $i$th nucleation pathway. For the
moderate degree of metastability of a macroscopic system, it is
expected that the critical size is independent on the time, i.e.
$n_c=n_c^{(i)}$. This is one of the basic postulates of the
classical nucleation theory~\cite{Kashchiev_Nucleation_2000},
which can be violated in the situation of the finite-sized system,
for
example~\cite{Mokshin/Barrat_JCP_2009,Tovbin_2012,Mokshin/Barrat_PRE_2008,Mokshin/Barrat_PRE_2010}.

(ii) The average waiting time of the $i$th nucleus with the
critical size, which appears in the system, is defined by
$\bar{\tau}_i(n_c)$. Consequently, the next relation should be
fulfilled: $\bar{\tau}_1(n_c) \leq \bar{\tau}_2(n_c) \leq
\bar{\tau}_3(n_c) \leq \ldots$, and the quantity $\tau_1 \equiv
\bar{\tau}_1(n_c)$ is the average waiting time of the first
(earliest) critical nucleus.

(iii) The time-dependent nucleation rate can be calculated
directly from the definition as
\begin{equation} \label{eq: NonSt_nucl_rate}
J(t)=\left . \frac{1}{V(t)}\frac{\partial\; i(t)}{\partial t}
\right |_{t=\bar{\tau}},
\end{equation}
where $i$ is the index number of the last nucleation event
appearing over the time scale $\bar{\tau}$. The quantity $i(t)$ in
Eq.~(\ref{eq: NonSt_nucl_rate}) is generated from
$\bar{\tau}_i(n_c)$ by serial accounting of the critical nuclei,
and, thereby, it indicates on the number of the overcritical
nuclei at the time $t = \bar{\tau}$. Note that Eq.~(\ref{eq:
NonSt_nucl_rate}) reproduces the most probable scenario for the
time-dependent nucleation rate $J(t)$ within the statistics of the
$M$ experiments.

(iv) The steady-state nucleation rate $J_s$ can be computed for
the linear part of the $i(t)$ plots according to Eq.~(\ref{eq:
NonSt_nucl_rate}). The time scale of the nucleation, which
precedes the regime of steady-state nucleation, is characterized
by the induction time $\tau_{ind}$. This quantity is determined as
the intersection point of the linear interpolation for $i(t)$ in
the regime of the steady-state nucleation,
\begin{equation} \label{eq: ind_time}
\frac{i(t)}{V(t)} = J_s(t-\tau_{ind}),
\end{equation}
with the time axis.

(v) The inverted MFPT curve $n_i(\bar{\tau})$, [where $\bar{\tau}$
is larger than the time scale of the $i$th critical nucleus,
$\bar{\tau}_{i}(n_c)$] will reproduce statistically the most
probable growth law of the $i$th nucleus. Consequently, such the
details of the nuclei growth, like the growth rate and the growth
exponent, appear to be available for the $i$th-order nucleus from
the extracted $n_i(\bar{\tau})$.

\textit{The Wedekind-Bartell scheme.}  -- The adjustment of the
MFPT concept to analyze the steady-state nucleation on the basis
of the MD simulations had been done and presented in two
works~\cite{Bartell_JCP_2006,Wedekind_JCP_2006} independently. The
authors have shown how  the steady-state nucleation rate $J_s$,
the critical size $n_c$, the Zeldovich factor $Z$ can be extracted
from the multiple simulations. The MFPT scheme of
Refs.~\cite{Bartell_JCP_2006,Wedekind_JCP_2006} -- the WB-scheme
-- suggests to utilize the temporal trajectories of the earliest
nucleus in the system and, thereby, it is focused on a single
quantity, $\bar{\tau}_1(n)$. According to the WB-scheme, the
steady-state nucleation rate is defined as
\begin{equation} \label{eq: MFPT_gauss}
J_s^{WB}  \propto \frac{1}{2\bar \tau_1(n_c)}.
\end{equation}
In fact, relation~(\ref{eq: MFPT_gauss}) evaluates the reduced
inverse average waiting time for the first critical nucleus, but
not the actual $J_s$. Nevertheless, the WB-scheme can provide the
correct values of $J_s$. This is possible in a very specific
situation, when the number of the critical nuclei increases with
time according to
\begin{equation}
i(t)^{WB} = H(t-\tau_1)\left ( \frac{t}{2\tau_1} + \frac{1}{2}
\right ),
\end{equation}
where $\tau_1 \equiv \bar{\tau}_1(n_c)$ and $H(\ldots)$ is the
Heaviside step function. It is also necessary to note that the
WB-scheme yields the negative induction time $\tau_{ind} = -
\tau_1$.
\begin{figure}[!htbp]
\includegraphics[width=8.4cm]{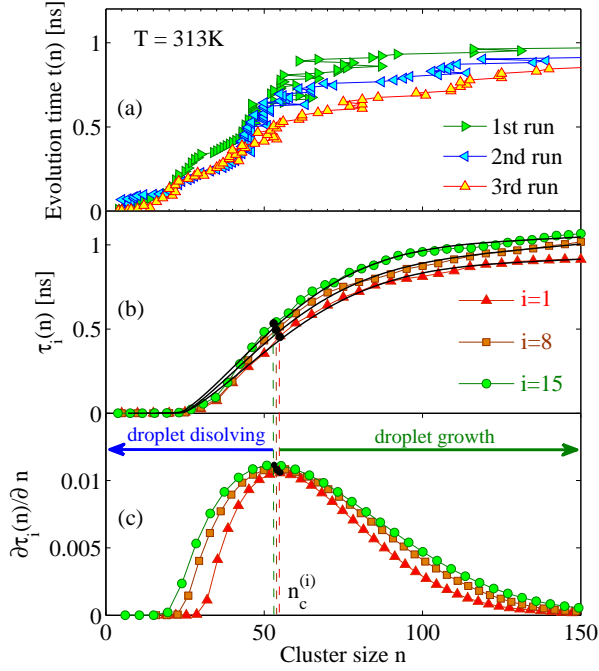}\vspace{-3 mm}
\caption{Analysis of the MD simulation data of water droplet
nucleation at the temperature $T=313$~K: (a) Inverted time
dependence of the largest droplet size $t(n^{(1)})$ as extracted
from three independent experiments. (b) MFPT curves for the $1$st,
$8$th and $15$th nucleated droplets defined from the set of
independent experiments by means of Eq.~(\ref{eq: MFPT_multiple}).
Trajectories presented in panel (a) were also utilized to define
the curve $\bar{\tau}_1(n)$. (c) First derivative $\partial
\bar{\tau}_i(n)/\partial n$ defined for the MFPT-curves given in
panel (b), $i=1$, $8$ and $15$. The position of the maximum on a
curve corresponds to the critical size $n_c^{(i)}$ of the $i$th
nucleated droplet. The average waiting time of the $i$th nucleus
can be extracted from the MFPT's as $\bar{\tau}_i(n=n_c^{(i)})$.
\label{fig: MFPT}} 
\end{figure}

\section{Droplet nucleation and growth in water vapor}

In this section we demonstrate the application of the numerical
scheme given in Sec.~\ref{Sec: MFPT} to characterize the droplet
nucleation and the subsequent droplet growth in water vapor on the
basis of the MD simulation data. These simulation data are
obtained within a way identical to given one in
Ref.~\cite{Mokshin/Galimzyanov_JPCB_2012}. Here, $N=8000$ water
molecules interact in the cubic cell effectively through the
anisotropic mW-potential, which represents the Stillinger-Weber
potential adopted for water
system~\cite{Molinero/Moore_JCPB_2009}. Therein, the periodic
boundary conditions in all directions were taken. The
consideration covers the temperatures from $T=273$~K to $353$~K at
the pressure $p\simeq 1$~atm, whilst the $NpT$-ensemble was
applied in the particular simulation runs. The samples at these
temperatures were prepared by means of the isobaric cooling from
the well equilibrated vapor configurations.~\footnote{Note that
the same cooling protocol was applied before to study ice
nucleation in supercooled liquid water and  to examine the
\emph{steady-state} droplet nucleation in water vapor in
Ref.~\cite{Molinero/Moore_JCPB_2009} and
Refs.~\cite{Matsubara_JCP_127_2007,Mokshin/Galimzyanov_JPCB_2012},
respectively.}

\begin{figure}[htbp]
\centering
\includegraphics[width=8.4cm]{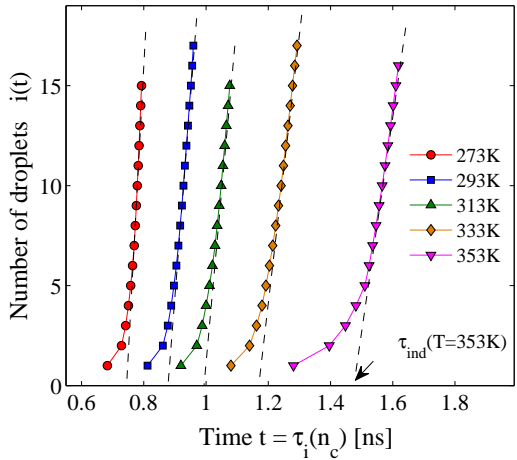}\vspace{-3 mm}
\caption{Isothermic time-dependent number of the nucleated
droplets. Each curve is resulted from the generated data for
$\bar{\tau}_i(n_c)$. Dash lines display the linear interpolation
of the data for the steady-state nucleation regime by
Eq.~(\ref{eq: ind_time}) with the induction time $\tau_{ind}$.
\label{fig: number_droplets}}
\end{figure}
\begin{table*}[ht]
  \caption{Rate characteristics of nucleation-growth kinetics of water
  droplets at
  temperature $T$ (K) and vapor number density $\rho_v$ ($\times 10^{-2}$\;nm$^{-3}$):
  average waiting time of the first droplet  $\bar{\tau}_1$ (ns),
  induction time $\tau_{ind}$ (ns),
  steady-sate nucleation rate $J_s$ ($\times 10^{32}$m$^{-3}$s$^{-1}$) defined according to Eq.~(\ref{eq: NonSt_nucl_rate}),
  steady-state nucleation rate $J_s^{WB}$ ($\times 10^{32}$m$^{-3}$s$^{-1}$) from Eq.~(\ref{eq: MFPT_gauss}),
  steady-state nucleation rate $J_s^{MG}$ ($\times 10^{32}$m$^{-3}$s$^{-1}$) taken from Ref.~\cite{Mokshin/Galimzyanov_JPCB_2012},
  and  growth factor $\mathcal{G}_{cl}$ (m$^{1/{\nu}}$s$^{-1}$) with the exponent $\nu=1.3$.
  }
   \label{Table: 1} 
  \begin{ruledtabular}
\bigskip
    \begin{tabular}{cccccccc}
 $T$ & $\rho_v$ & $\bar{\tau}_1$ & $\tau_{ind}$ & $J_s$ & $J_s^{WB}$ & $J_{s}^{MG}$  & $\mathcal{G}_{cl}$\\
    \hline
$273$ & $1.548$ & $0.68 \pm 0.05$ & $0.75 \pm 0.04$ & $84.6$ & $0.347$ & $0.345$ & $576 \pm 60$ \\
$293$ & $1.426$ & $0.81 \pm 0.07$ & $0.87 \pm 0.05$ & $45.6$ & $0.254$ & $0.263$ & $486 \pm 62$ \\
$313$ & $1.353$ & $0.92 \pm 0.08$ & $0.99 \pm 0.05$ & $34.1$ & $0.210$ & $0.211$ & $414 \pm 51$ \\
$333$ & $1.250$ & $1.08 \pm 0.11$ & $1.16 \pm 0.06$ & $23.7$ & $0.165$ & $0.165$ & $357 \pm 55$ \\
$353$ & $1.174$ & $1.28 \pm 0.13$ & $1.41 \pm 0.06$ & $18.6$ & $0.127$ & $0.126$ & $292 \pm 42$ \\
\end{tabular}
\end{ruledtabular}
\end{table*}
Thereupon, the samples of water vapor at the desired temperatures
$T=[273$, $293$, $313$, $333$, $353$]~K were ``equilibrated'' over
a time-scale $\sim 10$~ps, and the initial configurations for the
study of the phase transition kinetics were stored. We generated a
hundred independent configurations for each ($p,T$)-point to
perform the further statistical treatment within the method
presented in Sec.~\ref{Sec: MFPT}. Identification of the molecules
belonging to the vapor or the liquid phase was done using the
Stillinger
rule~\cite{Khusnutdinoff/Mokshin_set,Malenkov,Stillinger_JCP_1963}.
As a result, the time-dependent size $n_{\alpha i}(t)$ of each
liquid droplet for every simulation run was defined. We remind
that the term $\alpha \in [1,\;100]$ denotes the index of
simulation, and $i$ is the order number of the nucleated droplet
detected in the $\alpha$th simulation.

Figure~\ref{fig: MFPT}(a) shows the \emph{inverted} time-dependent
trajectories of the size, $\tau(n_{\alpha i})$, of the first
($i=1$) nucleated droplet as extracted from three different
simulations  ($\alpha=1$, $2$ and $3$) of the system at the
temperature $T=313$~K. Averaging over one hundred independent
trajectories (these three and others) according to Eq.~(\ref{eq:
MFPT_multiple}) yields the MFPT curve for the first nucleation
event, $\bar{\tau}_1(n)$, presented in Fig.~\ref{fig: MFPT}(b). In
addition, Fig.~\ref{fig: MFPT}(b) demonstrates the MFPT curves
computed for the $8$th and $15$th nucleated droplets. The MFPT
curves for other values of the index $i$ have the similar form,
and, therefore, are not shown. Following the scheme presented in
Sec.~\ref{Sec: MFPT}, the derivatives $\partial
\bar{\tau}_i(n)/\partial n$ were numerically calculated from the
MFPT's. As a result, the critical size $n_c^{(i)}$ for each $i$th
nucleation event as well as the corresponding nucleation time
scale $\bar{\tau}_i(n_c)$ were estimated from the positions of the
principal maximum in $\partial \bar{\tau}_i(n)/\partial n$ [see
Fig.~\ref{fig: MFPT}(c)]. As can be see from Fig.~\ref{fig:
MFPT}(c), the position of the maximum is unchanged with the order
of the nucleation event $i$. Consequently, the value of the
critical size $n_c$ remains unchanged during the nucleation
process in the system at the fixed temperature. We found that this
is valid for all the considered temperatures. Moreover, the
critical size decreases from $n_c\simeq 75$ molecules at the
temperature $T=273$~K to $n_c\simeq 42$ molecules at $T=353$~K
(see Ref.~\cite{Mokshin/Galimzyanov_JPCB_2012}).

Figure~\ref{fig: number_droplets} represents the time dependence
of the number of the nucleated water droplets in a system staying
at the isothermal-isobaric conditions. Since the time derivative
of this term, $\partial i(t)/\partial t$, defines the nucleation
rate $J(t,T)$ [see Eq.~(\ref{eq: NonSt_nucl_rate})], one can see
from Fig.~\ref{fig: number_droplets}, that the regime with
time-dependent nucleation rate $J(t,T)$ precedes the steady-state
nucleation, where the quantity $i(t)$ has a linear time dependence
and $J(t,T) \equiv J_s(T)$. The extracted values of the induction
time are given in Table~\ref{Table: 1}. The quantity
$\tau_{ind}(T)$ decreases with falling more deeply into the
metastable phase, i.e. with decrease of the temperature $T$, that
is in a qualitative agreement with the classical nucleation
theory~\cite{Kashchiev_Nucleation_2000}. The driving force of the
nucleation  grows with the temperature decreasing. Thereby, it
accelerates the nucleation and reduces the values of the induction
time. Moreover, the observed behavior of $\tau_{ind}(T)$ is
correlated with the temperature dependence of the waiting time
$\bar{\tau}_1(T)$. Namely, we find $\bar{\tau}_1(T)/\tau_{ind}(T)
= 0.92 \pm 0.01$ for the considered temperature range.

The steady-state nucleation rate $J_s(T)$ has been computed
according to Eq.~(\ref{eq: NonSt_nucl_rate}) as a slope of linear
part of each isothermic curve $i(t,T)$ shown in Fig.~\ref{fig:
number_droplets}. The extracted values of $J_s(T)$ are presented
in Table~\ref{Table: 1}, where they are compared with $J_s^{WB}$
obtained by means of the WB-scheme and with $J_s^{MG}$ derived in
Ref.~\cite{Mokshin/Galimzyanov_JPCB_2012} through the extended
WB-scheme. If the terms $J_s^{WB}$ and $J_s^{MG}$ have practically
the same values, then the both terms underestimate the values of
$J_s(T)$ by about two orders of magnitude. This difference
indicates that the WB-scheme with $J_s^{WB}$ as well as the
extended WB-scheme with $J_s^{MG}$ provides information about the
rate of the occurrence of the first (earliest) critical nucleus
and does not reproduce the actual steady-state nucleation rate
$J_s$ of the system. Remarkably, the values of $1/\bar{\tau}_1(T)$
are correlated with for the nucleation rate $J_s(T)$, albeit the
correlation is not linear.

\begin{figure}[!htbp]
\centering
\includegraphics[width=8.4cm]{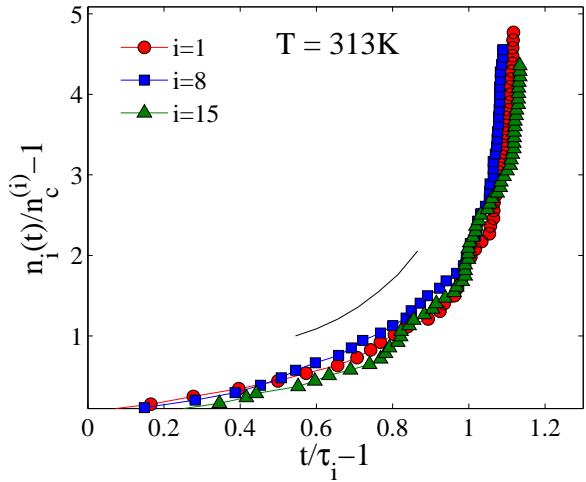} \vspace{-3 mm}
\caption{Rescaled growth curves of the water droplets with the
order number $i=1$, $8$ and $15$. The term $\tau_i$ is the waiting
time and $n_c^{(i)}$ is the critical size of the $i$th-order
nucleus. Note that the data are reproducible by Eq.~(\ref{eq:
growth_law}): solid curve shows fragment of the fitting shifted
upwards, for clarity. \label{fig: growth_curves}}
\end{figure}
The consistency of the computed values of $J_s$  with the known
experimental
data~\cite{Miller_JCP_1983,Viisanen_JCP_1993,Heath_JCP_2002,Khan_JCP_2003,Klein_JPCA_2005}
can be verified by plotting $\log J_s$ versus the scaled quantity
$(T_c/T -1 )^3/[\ln(p/p_{eq})]^2$, where a straight line should
appear. The resulting graph is known as the scaled Hale
plot~\cite{Hale_PRL_2010,Hale_PRA_1986}. If the construction of
the Hale plot is explicit with the experimental data, then for the
treatment of the MD simulation results one needs to know the exact
values of the critical temperature $T_c$ and the equilibrium vapor
pressure $p_{eq}$ as the quantities generated by the concrete
model of potential. The critical temperature and the saturation
curve are not determined for the mW-model; therefore, the
corresponding scaled constructions can be
speculative~\cite{Zipoli}. The direct comparison reveals that the
values of $J_s$ obtained within the mW-model overestimate all the
known experimental data by the orders of magnitude, but they
approach the result of the MD simulations performed by Yasuoka and
Matsumoto with the TIP$4$P-model: $J_s=9.62 \times
10^{32}$\;m$^{-3}$s$^{-1}$ at $T=350\;K$ and $\rho_v = 1.55 \times
10^{-2}$\;nm$^{-3}$~~~~~\cite{Yasuoka/Matsumoto_JCP_1998,Merikanto_JCP_2004}.

The correctness of the extracted values of the nucleation rate
$J(t)$ can be confirmed if these data at the known droplet growth
law will reproduce the MD simulation results for the overall
condensation kinetics~\cite{Mokshin/Barrat_PRE_2010}. The growth
of the earliest ($i=1$) droplet for the system can be taken as
\begin{equation} \label{eq: power_law_radius}
R(t)=R_c + \left ( \mathcal{G}_{cl} t \right )^{\nu},
\end{equation}
where $R$ is the droplet radius at the time $t$ elapsed after the
appearance of the critical droplet with the radius
$R_c$~\cite{Mokshin/Galimzyanov}. The values of the growth factor
$\mathcal{G}_{cl}$ are given in Table~\ref{Table: 1}; and the
growth exponent $\nu=1.3$ appears to be the
temperature-independent term. The last equation can be rewritten
in the rescaled form for the arbitrary $i$th-order nucleated
droplet:
\begin{equation} \label{eq: growth_law}
\frac{n_i(t,\bar{\tau}_i)}{n_c^{(i)}} = \left [ 1 + \frac{
\mathcal{G}_{cl}^{\nu} \bar{\tau}_i^{\nu}}{R_c^{(i)}} \left (
\frac{t}{\bar{\tau}_i} -1 \right )^{\nu} \right ]^3,
\end{equation}
where $n(t) = c_g \rho_c R(t)^3$, $\rho_c$ is the number density
of the nucleated phase and $c_g$ is the shape-factor, which is
$c_g = 4\pi/3$ for a sphere. It is usually accepted by the
theoretical treatments that the growth of all the overcritical
nuclei ($i=1$, $2$, $\ldots$) in the system follows the same
growth law~\cite{Schneidman_JCP_1994,Kalikmanov_JCP_2008}. The
scheme presented in Sec.~\ref{Sec: MFPT} allows one to clarify how
the growth of the nuclei emergent at the different times proceeds.
Figure~\ref{fig: growth_curves} presents the rescaled growth
curves of the nuclei with the order number $i=1$, $8$ and $15$ for
the system at the temperature $T = 313$~K. Here, each curve is
obtained by means of inverted averaging of the independent
pathways $n_{\alpha i}(t)$ [see item (v) in Sec.~\ref{Sec: MFPT}].
As can be seen, the droplets with the different order numbers
($i=1$, $8$ and $15$) evolve according to the same growth
trajectory, which is reproducible by Eq.~(\ref{eq: growth_law}).
Moreover, we found that this result is valid for the droplet
growth at other considered temperatures.

\begin{figure}[!htbp] \centering
\includegraphics[width=7.5cm]{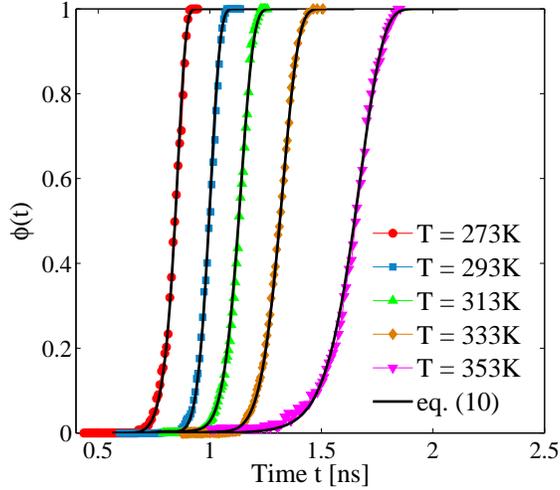} \vspace{-3 mm}
\caption{Time evolution of the condensed liquid fraction in the
water vapor at the different temperatures. Markers present results
obtained from simulations. Solid curves are predictions of
Eq.~(\ref{eq: KJMA_simple}), which contains no adjustable
parameters. \label{fig: fraction}}
\end{figure}
The vapor condensation as an overall phase transition is defined
by the droplet nucleation and growth. Therefore, the fraction of
material condensed into a liquid phase $\phi(t)$ evolves with the
time according to
\begin{equation} \label{eq: KJMA_eq}
\phi(t) =  1 - \exp\left \{ - c_g \int_{0}^t J(t') \left [
\int_{t'}^t G(t'') dt''  \right ]^3 dt' \right \} ,
\end{equation}
that is the main equation of the KJMA
theory~\cite{Kashchiev_Nucleation_2000}. Here, $G(t) = dR(t)/dt$
is the instantaneous growth rate of the droplet radius. Taking
into account that the critical size is unchangeable with the time
and $c_g \simeq 4\pi/3$, one obtains
\begin{equation} \label{eq: KJMA_simple}
\phi(t) =  1 - \exp\left \{ - \frac{4\pi R_c^3}{3}  \int_{0}^t
J(t') \left [ 1 + \frac{\mathcal{G}_{cl}^{\nu}}{R_c} \left (t -
t'\right )^{\nu} \right ]^3 dt' \right \} .
\end{equation}
Although a single $\alpha$th experiment produces the inherent
trajectory $\phi_{\alpha}(t)$, the most probable scenario of the
condensation kinetics $\phi(t)$ can be restored on the basis of
the set $\phi_1(t)$, $\phi_2(t)$, $\ldots$, $\phi_{\alpha}(t)$,
$\ldots$, $\phi_M(t)$ through the inverted averaging [see
Eq.~(\ref{eq: MFPT_single})]~\footnote{Here, the variable
$t=\bar{\tau}$ is the mean first-passage time for the quantity
$\phi$.}. Thus, the fraction $\phi(t)$ is a measurable term. On
the other hand, the time-dependent nucleation rate $J(t)$, the
growth factor $\mathcal{G}_{cl}$ and the exponent $\nu$ are
computed via the scheme presented in Sec.~\ref{Sec: MFPT} and,
therefore, can be used as input parameters in Eq.~(\ref{eq:
KJMA_simple}). As seen from Fig.~\ref{fig: fraction},  remarkable
agreement is observed between the measured data for $\phi(t)$ and
the predictions of Eq.~(\ref{eq: KJMA_simple}). Importantly, the
agreement is observed for all the isotherms, and no fitting
parameters are utilized within Eq.~(\ref{eq: KJMA_simple}). This
is evidence that the computational method presented here provides
the precise estimates for the nucleation-growth rates. Thereby, it
can be used as a convenient tool to study the overall phase
transition kinetics.

\section{Conclusions}

The MFPT approach is known as a pure theoretical concept, which
can be applied, for example, for the study of the barrier-crossing
problems~\cite{Yvinec_JCP_2012,Hanggi_RMP_1990}. The application
of this approach for analyzing the nucleation kinetics within the
MD simulation data had been demonstrated by Wedekind \textit{et
al.}~\cite{Wedekind_JCP_2006} and Bartell \textit{et
al.}~\cite{Bartell_JCP_2006}. Recently, the extended version of
the approach was reported in
Ref.~\cite{Mokshin/Galimzyanov_JPCB_2012}. There are two notable
points, which define the applicability of the MFPT approach in
such realization. First, the WB-scheme (in a version of
Refs.~\cite{Bartell_JCP_2006,Wedekind_JCP_2006}) is restricted by
consideration of the \textit{steady-state} nucleation, where the
nucleation rate $J_s$ has no dependence on the time. This
corresponds to the distinct temporal regime of the steady-state
nucleation within a phase transition passage. Second, the
WB-scheme assumes that the waiting time for the first nucleation
event, $\bar{\tau}_1(n_c)$, and the steady-state nucleation time
scale, $\tau_s = (J_s V)^{-1}$ are related as $\tau_s = 2
\bar{\tau}_1(n_c)$, that is appropriate for a very specific case
of nucleation kinetics. In particular, this condition can be not
relevant for the non-equilibrium or driven phase
transitions~\cite{Mokshin/Galimzyanov/Barrat_PRE_2013}.

The method presented in this work allows one to define the most
probable time scales of the successive nucleation events within a
statistics of the independent experiments. Thereby, the nucleation
rate can be estimated directly according to its definition as a
time-dependent term. Therewith, the critical size and the growth
trajectories of the emergent nuclei can be restored with a
precision.

In this work, we have analyzed the overall vapor-to-liquid
transition kinetics of water within the presented statistical
method as applied to the coarse-grained MD simulation data. It is
notable that the induction time $\tau_{ind}(T)$ is correlated with
the waiting time for the first nucleation event $\bar{\tau}_1(T)$,
while both quantities, $\tau_{ind}(T)$ and $\bar{\tau}_1(T)$,
exceed the time scale of the steady-state nucleation $\tau_s(T)$
more than by two orders in magnitude. In addition, no time-
($i$-th order) dependence is detected in the values of the
critical size as well as in the character of the droplet growth,
that is in agreement with the classical nucleation theory.
Finally, we found that the extracted values of the
nucleation-growth rates are capable within the KJMA theory to
reproduce correctly the MD simulation results for the transformed
fraction $\phi(t)$.

\begin{acknowledgments}
The authors acknowledge R.~M.~Khusnutdinoff, Yu.~K.~Tovbin,
G.~G.~Malenkov, V.~V.~Brazhkin and V.~N.~Ryzhov  for motivating
discussions and their interest to our work.
B.N.G. was supported by the Russian Foundation for
Basic Research (project No.~$12$-$02$-$31228$).
\end{acknowledgments}

\bibliographystyle{unsrt}

\end{document}